\documentclass{aa}
\usepackage{epsfig}
\usepackage{graphicx,natbib}

\newcommand{\mr}{\mathrm}

\begin{document}

\title{Electron acceleration during three-dimensional relaxation of an
electron beam-return current plasma system in a magnetic field}

\titlerunning{Electron acceleration during three-dimensional beam-plasma relaxation}
\authorrunning{Karlick\'y and Kontar}

\author{M. Karlick\'y \inst{1}  \and E. P. Kontar\inst{2}}

\offprints{M.~Karlick\'y, \email{karlicky@asu.cas.cz}}

\institute{\inst{1}Astronomical Institute of the Academy of Sciences of the Czech
Republic, 25165 Ond\v{r}ejov, Czech Republic, e-mail: karlicky@asu.cas.cz\\
\inst{2}SUPA, School of Physics \& Astronomy, University of Glasgow, G12 8QQ, United Kingdom}

\date{Received  / Accepted }

  \abstract
   {}
  {We investigate the effects of acceleration during non-linear electron-beam
   relaxation in magnetized plasma
   in the case of electron transport in solar flares.}
   {The evolution of electron distribution functions is computed using a
   three-dimensional particle-in-cell electromagnetic code.
   Analytical estimations under simplified assumptions are made to provide comparisons.}
{We show that, during the non-linear evolution of the beam-plasma system, the
accelerated electron population appears. We found that, although the electron
beam loses its energy efficiently to the thermal plasma, a noticeable part of
the electron population is accelerated. For model cases with initially
monoenergetic beams in uniform plasma, we found that the amount of energy in
the accelerated electrons above the injected beam-electron energy varies
depending the plasma conditions and could be around 10-30\% of the initial beam
energy.}
 {This type of acceleration could be important for the interpretation of non-thermal
 electron populations in solar flares. Its neglect could lead to the over-estimation of
 accelerated electron numbers.
 The results emphasize that collective plasma effects should not be
 treated simply as an additional
 energy-loss mechanism, when hard X-ray emission in solar flares is interpreted,
 notably in the case of \textit{RHESSI} data.}

 \keywords{Sun: flares -- Sun: particle emission -- Sun: X-rays, gamma rays}

\maketitle

\section{Introduction}

Solar flare X-ray observations provide often unique insights into the processes
of electron acceleration and transport. Recent observations of solar flares,
notably with {\it RHESSI}  \citep{2002SoPh..210....3L} have emphasized the high
efficiency of electron acceleration in solar flares  [for a recent review of
electron properties, see \citet{2011SSRv..159..301K} and for the corresponding
implications for particle transport in solar flares,
\citet{1997SSRv...81..143K} and \citet{2011SSRv..159..107H}].

The high efficiency of electron acceleration results in high electron fluxes or
concentrations of deka-keV electrons in solar flares, and the subsequent
importance of collective effects to particle transport in the solar atmosphere.
The presence of large number of energetic electrons in coronal loops could
trigger a number of instabilities and generate plasma waves, which in turn
affect the transport of energetic particles from the acceleration region down
to the chromosphere. It has been shown that accounting for these collective
effects could affect the interpretation of hard X-ray spectra and lead to
additional observational consequences that are essential to the study of solar
flares. For example, the inclusion of Langmuir wave generation in the treatment
of spatially localized electron beams \citep{2009ApJ...707L..45H} prevents the
formation of a pronounced low-energy cut-off that appears in purely collisional
models \citep[e.g.][]{1971SoPh...18..489B,2002SoPh..210..373B}. Weibel
instability \citep{1959PhRvL...2...83W}  can quickly increase the velocities in
the direction perpendicular to the beam propagation and affect the observed
X-ray anisotropy \citep{2009NPGeo..16..525K,2009A&A...506.1437K}. The presence
of Langmuir waves in a flaring loop could also result in plasma emission
\citep[e.g.][]{1979ApJ...233..717V,1984ApJ...279..882E,1987ApJ...321..721H},
providing additional constraints on non-thermal electron populations in solar
flares.

In laboratory plasma experiments, collision-less effects involving various
instabilities have been demonstrated to play a key role in electron transport.
In experiments to study the beam-plasma interaction, the appearance of
electrons with energies exceeding that of the injected beam energy was noted in
early studies
\citep[e.g.][]{1964JNuE....6..173B,1968CzJPh..18..652F,1983SvPhU..26..116K}.
These above-the-injected energy electrons are normally connected to either the
presence of plasma inhomogeneities
\citep{1967PlPh....9..719V,1969JETP...30..131R,1976JPSJ...41.1757N,1979PhFl...22..321E}
or the nonlinear effects of wave-particle interactions
\citep{1975JETPL..21..192B,2010PhPl...17h3111T}. Furthermore, the influence of
plasma inhomogeneities or wave-wave interactions on beam-plasma instability in
the solar context has been studied extensively in connection to the theory of
solar radio-type III bursts
\citep[e.g.][]{1985SoPh...96..181M,2007PhPl...14l2111T,2009ApJ...695L.140K,2010SoPh..267..393T,2010ApJ...721..864R,2011ApJ...727...16Z}.
The processes governing non-thermal electron evolution in solar flares span a
computationally prohibitive range of timescales from the inverse plasma period
$\sim \omega_{pe}^{-1}$ to observational timescales of thousands of seconds.
Therefore, modeling efforts in solar flare physics have focused either on
relatively long observational timescales by either ignoring short-timescales or
addressing the micro-physics on scales smaller than the observational scales
\citep[e.g.][]{2009ApJ...699..990B,2001JPlPh..68..161B,2006NJPh....8...55E,1991AdSpR..11..193H,2008A&A...478..889L,2002A&A...382..301M,2002PhPl....9.1000R,2006A&A...457..313S,2004PhPl...11.5547S}.

\citet{2012A&A...539A..43K} showed that the $k$-spectrum evolution of
beam-generated Langmuir waves in collisional plasma could result in substantial
energy gain by high energy electrons or the effective acceleration of electrons
of the same beam above $~20$~keV for solar flare conditions. Thus, if
collisional relaxation is assumed, the number of energetic electrons inferred
from X-ray spectra could be overestimated and may lead to an apparently large
number of accelerated electrons. However, the treatment of
\citet{2012A&A...539A..43K} is a one-dimensional analysis based on weak
turbulence theory. They showed that  a positive density gradient, externally
excited density fluctuations, and the three-wave interaction involving
ion-sound mode could lead to an effective acceleration of beam electrons.
However, the role of the guiding magnetic field as well as three-dimensional
(3D) aspects of beam-plasma interaction have not yet been addressed.

This paper investigates the non-linear evolution of non-thermal electron beams
in a plasma. Using the particle-in-cell (PIC) model developed in
\citet{2009ApJ...690..189K} and \citet{2009A&A...506.1437K}, we investigate the
effects of the acceleration of electrons in the beam during a non-linear stage
of beam-plasma instability for various values of  the guiding magnetic field.
We show that about $\simeq 10-30\%$ of electrons are accelerated to the
energies greater than the energies at which they were injected. The presence of
a guiding field increases the number of electrons accelerated during a
beam-plasma interaction. The results show the appearance of accelerated
electrons and highlight that collective effects can lead to not only additional
energy losses for lower energy electrons but the acceleration of some electrons
above the initial energy of electrons.  As the hard X-ray spectra in solar
flares is normally due to deka-keV electrons, this acceleration effect should
be taken into account when the number of accelerated electrons is estimated. In
this paper, we compared these results with analytical estimates.


\begin{figure*}[!t]
  \begin{center}
    \epsfig{file=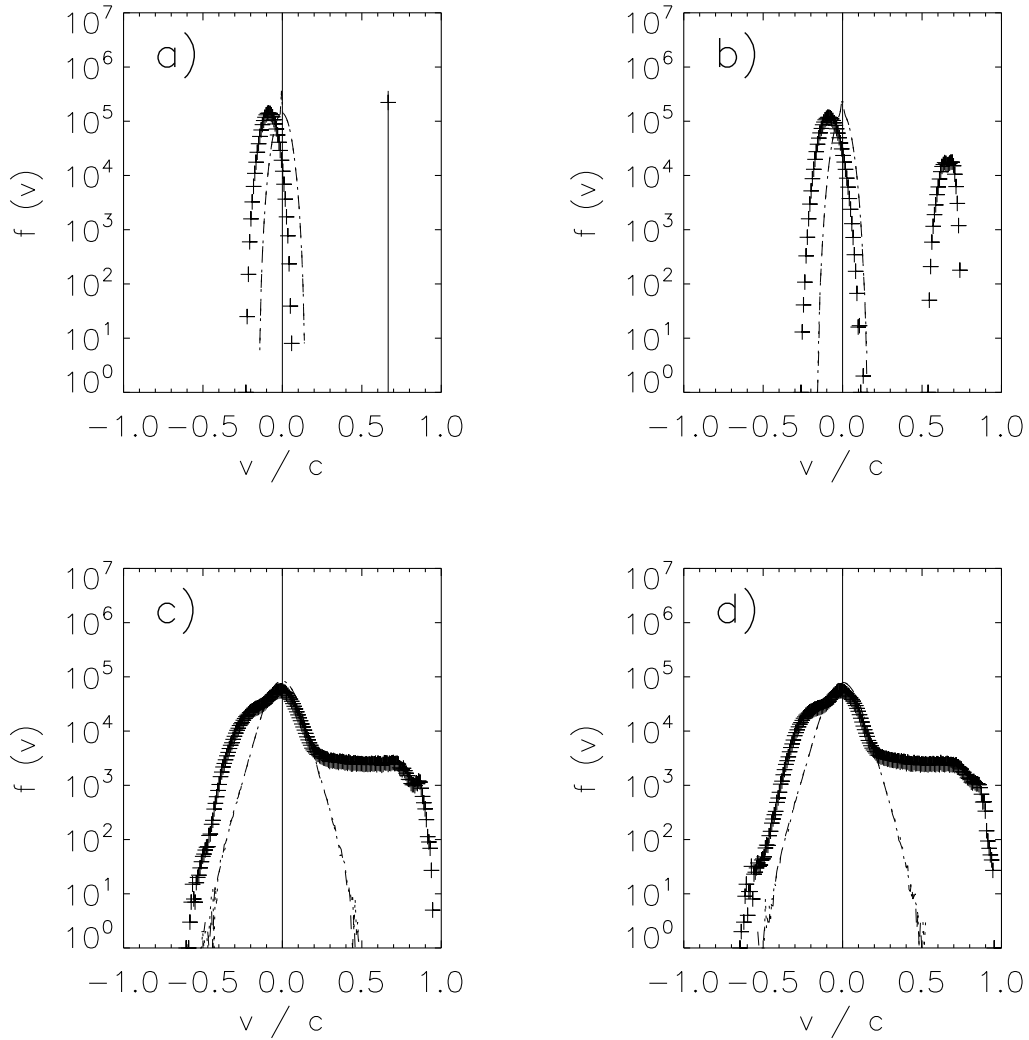, width=0.8\textwidth}
        \end{center}
    \caption{The electron velocity distributions for model E
    at four different times:
    at the initial state
 (a), at $\omega_\mathrm{pe} t$ = 40 (b), at  $\omega_\mathrm{pe} t$ = 140
 (c), and $\omega_\mathrm{pe} t$ = 200 (d).
 Crosses correspond to $f(v_{z})$, dotted and dashed lines display $f(v_x)$ and
$f(v_y)$, respectively.
  Note that $f(v_x)$ and $f(v_y)$ overlap.
  The vertical line in the part a) at $v/c$ = 0.666
  denotes the monoenergetic electron beam.}
  \label{fig1}
\end{figure*}

\begin{figure*}[!t]
  \begin{center}
    \epsfig{file=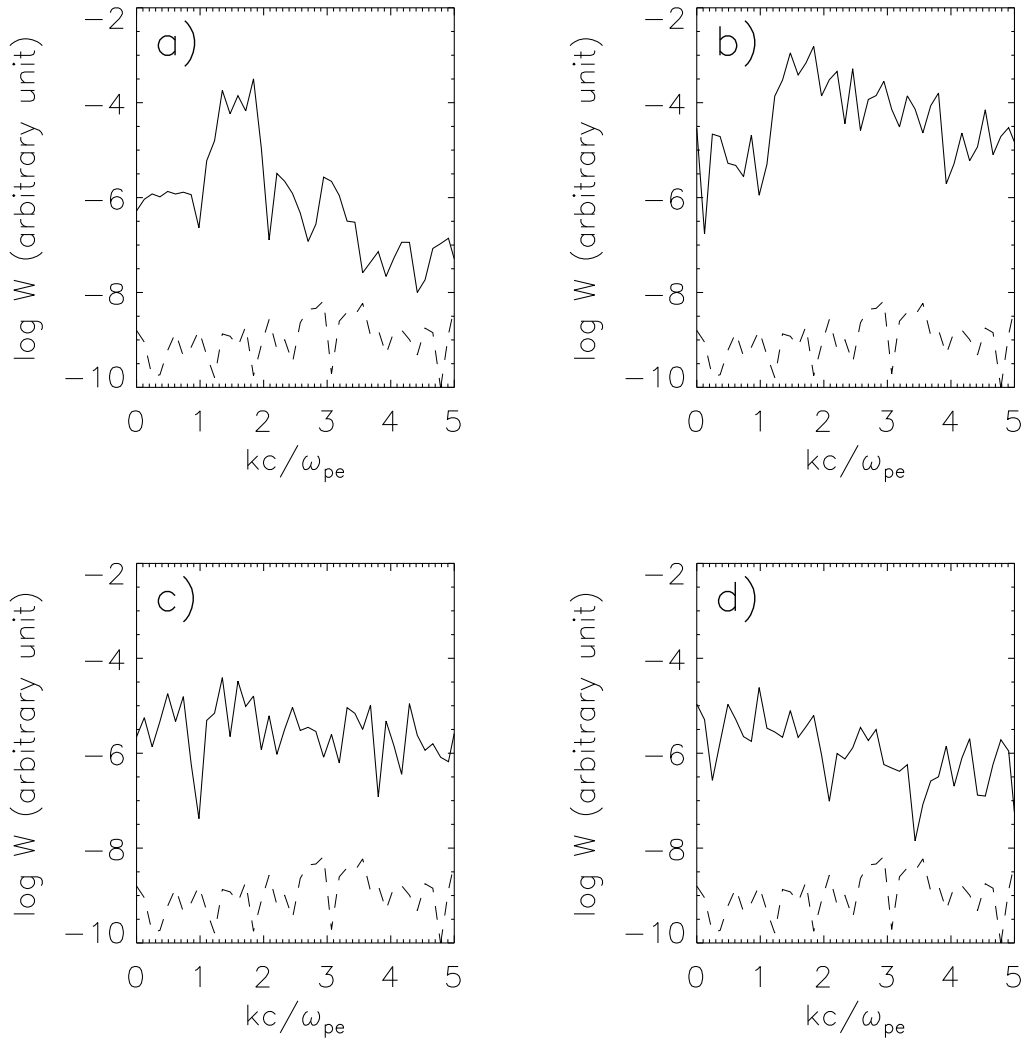, width=0.8\textwidth}
        \end{center}
    \caption{The Langmuir wave energy for model E in the $k$-space at four different times:
    $\omega_\mathrm{pe} t$ = 40
 (a), $\omega_\mathrm{pe} t$ = 60 (b), $\omega_\mathrm{pe} t$ = 140
 (c), and $\omega_\mathrm{pe} t$ = 200 (d) (solid lines).
 For comparison in each panel, the initial Langmuir wave energy is
added (dashed line).}
  \label{fig2}
\end{figure*}

\begin{figure*}[!t]
  \begin{center}
    \epsfig{file=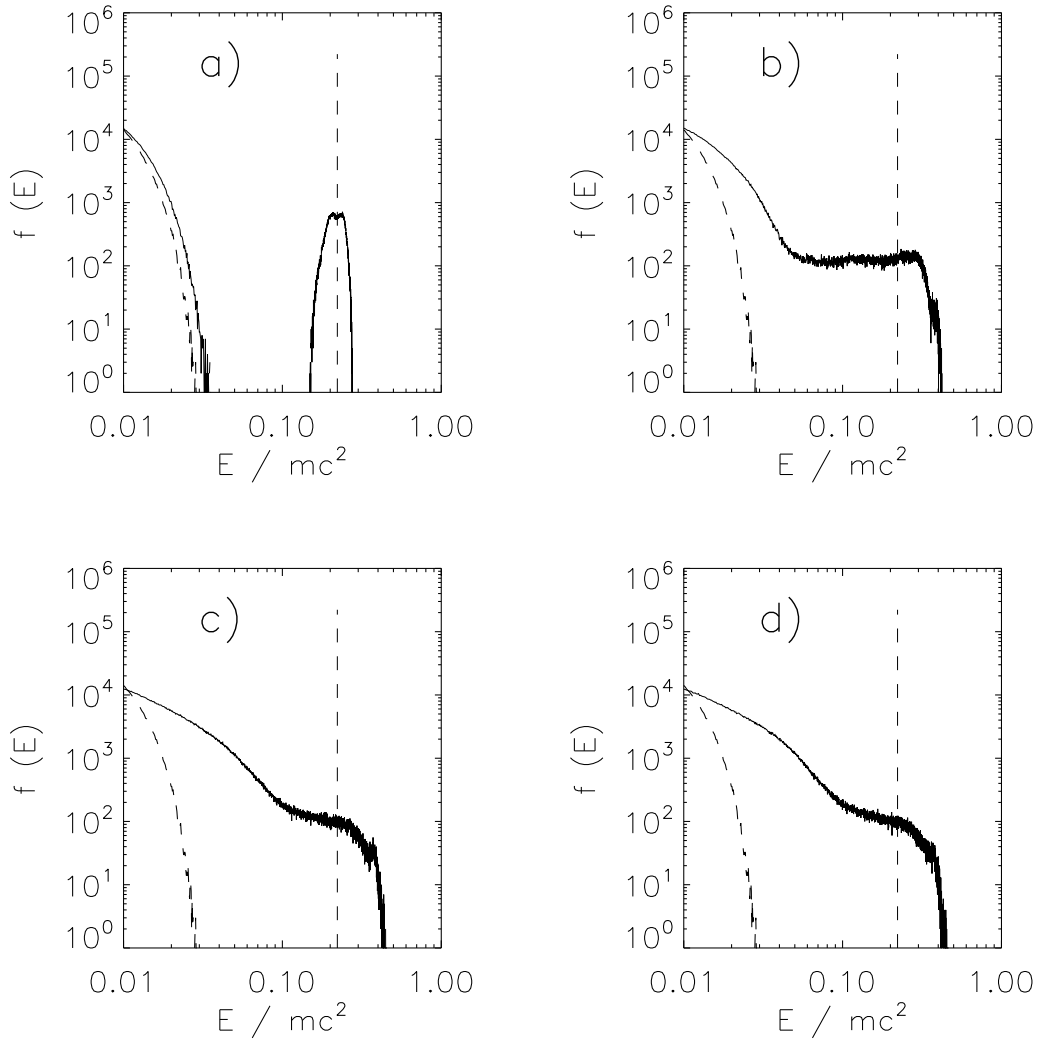, width=0.8\textwidth}
        \end{center}
    \caption{The electron energy distributions for model E
    at four different times:
    $\omega_\mathrm{pe} t$ = 40
 (a), $\omega_\mathrm{pe} t$ = 60 (b), $\omega_\mathrm{pe} t$ = 140
 (c), and $\omega_\mathrm{pe} t$ = 200 (d) (solid lines).
For comparison in each panel, the initial electron-plasma distribution together
with the initial monoenergetic beam are added (dashed lines).}
  \label{fig3}
\end{figure*}

\begin{figure*}[t]
  \begin{center}
    \epsfig{file=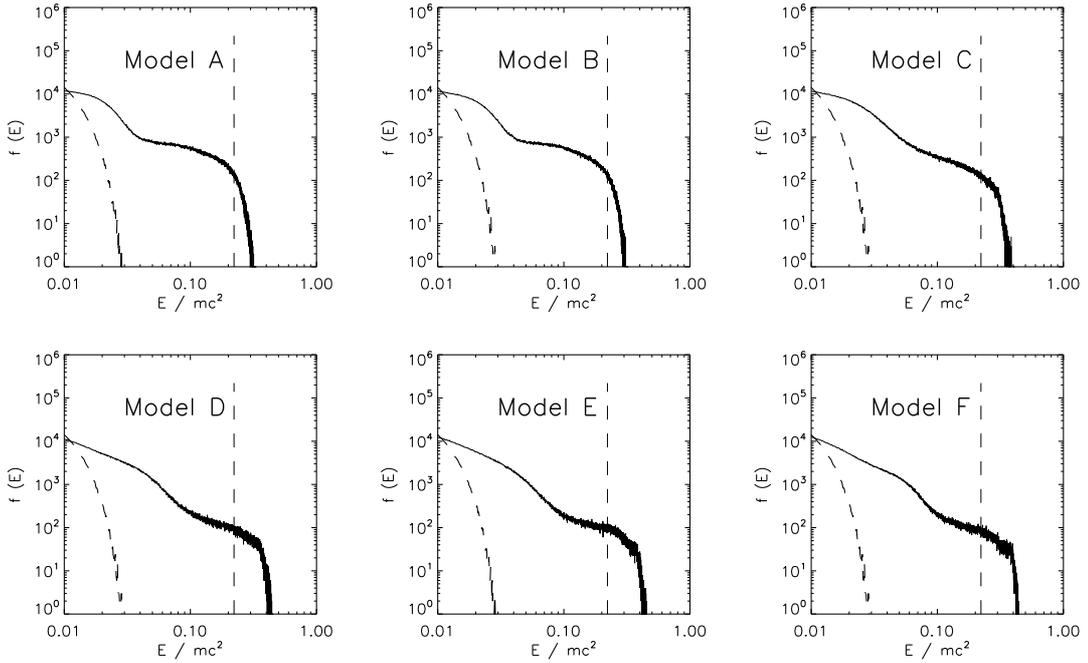, width=0.8\textwidth}
        \end{center}
    \caption{The electron energy distributions (solid lines)
    at $\omega_\mathrm{pe} t$ = 200
    as a function of the magnetic field
    in models A-F with $\omega_\mathrm{ce}/\omega_\mathrm{pe}$ =
    0.0, 0.1, 0.5, 0.7, 1.0, and 1.3, respectively. For comparison
    in each panel we plot the initial electron plasma distribution together with
    the initial monoenergetic beam (dashed lines).}
  \label{fig4}
\end{figure*}

\section{Simulation model}

\begin{table}[t]
\begin{minipage}[t]{\columnwidth}
\caption{Model parameters and the fraction (FR) of the beam energy
in the electrons with energies greater than the initial-beam electron energy.}
\label{tab4}
\centering
\renewcommand{\footnoterule}{}
\begin{tabular}{cccccc}
\hline
Model & $m_\mathrm{i}$/$m_\mathrm{e}$  & $n_\mathrm{b}$/$n_\mathrm{e}$  & $v_\mathrm{b}/c$ & $\omega_\mathrm{ce}/\omega_\mathrm{pe}$ & FR ($\%$)\\
\hline
A & 16 & 1/8 & 0.666 & 0.0 & 10 \\
B & 16 & 1/8 & 0.666 & 0.1 & 10 \\
C & 16 & 1/8 & 0.666 & 0.5 & 22 \\
D & 16 & 1/8 & 0.666 & 0.7 & 28 \\
E & 16 & 1/8 & 0.666 & 1.0 & 29 \\
F & 16 & 1/8 & 0.666 & 1.3 & 27 \\
G & 16 & 1/40 & 0.666 & 0.0 & 12 \\
H & 16 & 1/40 & 0.666 & 1.0 & 28 \\
\hline
\end{tabular}
\end{minipage}
\end{table}

It is commonly accepted that electrons in solar flares are accelerated at low
corona heights by primary energy-release processes. They propagate along the
magnetic field lines as electron beams downwards to loop footpoints, where they
generate hard X-ray emission. The rapid variations (of timescales $\sim$ 45
msec) in this hard X-ray emission observed in some events indicate that the
acceleration as well as the electron beam flux can may themselves rapidly
varying \citep[e.g.][]{1983ApJ...265L..99K,1995ApJ...455..347A}. This means
that the effect of fast propagation needs to be considered in the evolution of
these electron beams. Assuming that the electrons have a power-law distribution
in the acceleration region.  Owing to the propagation, fast electrons could
then overtake slower ones at some distance from the acceleration site
\citep[e.g.][]{2009ApJ...707L..45H}, forming an unstable distribution that can
be approximated by the mono-energetic beam.

For our study, we used a 3D (three spatial and three velocity components)
relativistic electromagnetic PIC code \citep{2009ApJ...690..189K}. The system
sizes are $L_x$ = 45$\Delta$, $L_y$ = 45$\Delta$, and $L_z$ = 600$\Delta$,
where $\Delta$ is the grid size.

We initiated a spatially homogeneous electron-proton plasma with the
proton-electron mass ratio $m_\mathrm{p}/m_\mathrm{e}$=16 (models A-H in Table
1). This ratio was chosen to shorten the computational times and keep the
proton skin-depth shorter than the dimensions of the numerical box.
Nevertheless, the ratio was still sufficient to clearly separate the dynamics
of electrons and protons. The electron thermal velocity was $v_{T\mathrm{e}}$ =
0.06 $c$, where $c$ is the speed of light. In all models, 160 electrons and 160
protons per cube grid were used. The electron plasma frequency was
$\omega_\mathrm{pe}$ = 0.05 $(\Delta t)^{-1}$ ($\Delta t$ = 1 is the time step)
and the electron Debye length was $\lambda_\mathrm{D}$ =
$v_{T\mathrm{e}}$/$\omega_\mathrm{pe}$ = 0.6 $\Delta$. The electron and proton
skin-depths were $\lambda_\mathrm{ce}$ = $c/\omega_\mathrm{pe}$ = 10 $\Delta$
and $\lambda_\mathrm{cp}$ = $c/\omega_\mathrm{pp}$ = 40 $\Delta$ (where
$\omega_\mathrm{pp}$ is the proton plasma frequency), respectively.

We then included a mono-energetic beam that was homogeneous throughout the
numerical box. To keep the total current zero in these models in their initial
states, we introduced an initial return current by shifting the background
plasma electrons in the velocity space according to the relation $v_\mathrm{d}
= - v_\mathrm{b} n_\mathrm{b}/n_\mathrm{e}$, where $v_\mathrm{b}$ is the
velocity of the electron beam, and $n_\mathrm{b}$ and $n_\mathrm{e}$ are the
beam and background plasma densities; for an example of this type of
initialization, see \citet{2008ApJ...684.1174N}. In principle, it is possible
to start from a zero initial return current. However, owing to inductive
effects included in the used 3-D electromagnetic code, the current starts to
oscillate at the electron plasma frequency with the amplitude of the
stabilizing return current. Therefore in these simulations, we recommend
starting from an already stabilized return current \citep{2008N}. Although
these current oscillations can influence the studied acceleration, for
simplification, we did not take this effect into account. We similarly
neglected any effects of the return-current formation (electrostatic ones) on
the front of the propagating beams.

The beam velocity was chosen to be $v_\mathrm{b}/c =  0.666$  in the $z$
direction. The ratio of the beam to the plasma densities was chosen to show the
details of the acceleration process $n_\mathrm{b}/n_\mathrm{e}= 1/8$, which is
a rather high value (however still realistic). The lower values of
$n_\mathrm{b}/n_\mathrm{e}= 1/40$ , see Table 1, for comparison, were used  in
models G and H.

To investigate the influence of the magnetic field in the models, we consider
several values of the background magnetic field, so that the ratio of the
electron-cyclotron to electron-plasma frequencies
($\omega_\mathrm{ce}/\omega_\mathrm{pe} $) is 0.0, 0.1, 0.5, 0.7, 1.0, and 1.3
(see Table 1). In all models, the periodic boundary conditions were used.

\section{Results of numerical simulations}

Using the above described model, we performed eight runs (A-H) using the
parameters given in Table 1. As an example (see Fig.~\ref{fig1}), a time
evolution of the electron velocity distribution for model E is shown at four
different times: the initial state (a), $\omega_\mathrm{pe} t$ = 40 (b),
 $\omega_\mathrm{pe} t$ = 140 (c),  and $\omega_\mathrm{pe} t$ = 200 (d).
Crosses correspond to $f(v_{z})$, and dotted and dashed lines display $f(v_x)$
and $f(v_y)$, respectively.  The vertical line in the initial state at $v/c$ =
0.666 denotes the mono-energetic electron beam. As can be seen here, a plateau
forms in the velocity space between the velocity of the initial beam and that
of the background plasma electrons. It is produced by the so-called
quasi-linear relaxation, in which the electron beam generates the Langmuir
waves (see Fig.~\ref{fig2}). The Langmuir waves were initially generated at the
$k$-wave vector corresponding to the resonance condition for the beam-plasma
instability, i.e. at $\omega$ = $k_z v_b$ (see the peak in the Langmuir wave
energy-spectrum in Fig.~\ref{fig2}a), where $\omega$ is the frequency of the
Langmuir waves and $v_\mathrm{b}$ is the electron beam velocity
\citep{1975MAtom....R....M}. These Langmuir waves then decay and merge mainly
in accordance with the three wave interactions (for details, see e.g.
\citet{2000A&A...353..757B}) and evolve in the $k$-vector space (see the
enhanced Langmuir wave spectrum in Fig.~\ref{fig2}b,c,d). The Langmuir waves,
on the other hand, scatter the beam electrons and heat the background plasma.
During this process, the plateau in the electron distribution function
$f(v_{z})$ is formed; (see detailed discussion of the quasi-linear relaxation
in the following section). However, during this process some beam electrons,
owing to an interaction with the Langmuir waves, obtain an energy that is
higher than the electron energy in the initial beam. This is shown in
Fig.~\ref{fig3}, where a time evolution of the electron energy distributions
for model E  at four different times is presented: at (a) $\omega_\mathrm{pe}
t= 40$ (b) $\omega_\mathrm{pe} t$ = 60 , (c) $\omega_\mathrm{pe} t$ = 140, and
(d) $\omega_\mathrm{pe} t$ = 200. For comparison, we show in each panel in this
figure the initial electron-plasma distribution together with the initial
mono-energetic beam (dashed lines). Here, it can clearly be seen that some
electrons have the energies higher than those of the initial beam.

\begin{figure}[!t]
  \begin{center}
    \epsfig{file=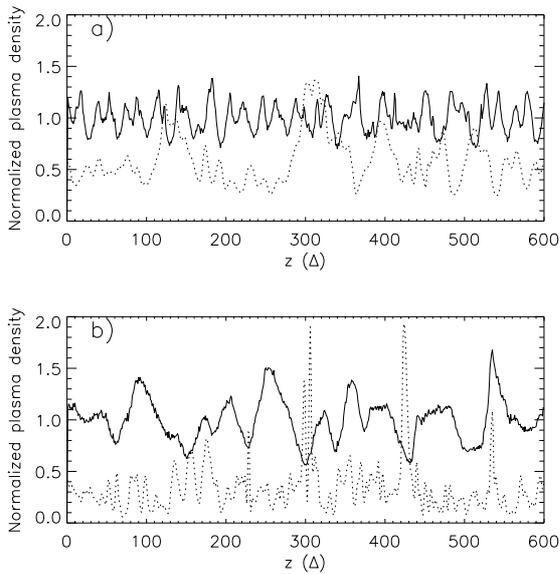, width=8cm}
        \end{center}
    \caption{Normalized plasma density (solid line) and the
    electric field density (dashed line) in the $z$-direction along the line with
    $x$ = $y$ = 22$\Delta$ for model E
    at two times: $\omega_\mathrm{pe} t$ = 60
 (a) and $\omega_\mathrm{pe} t$ = 190 (b). The electric field density is expressed in
 arbitrary units, but the electric field density at (b) is multiplied by a factor of 50.}
  \label{fig5}
\end{figure}

Although in the initial state, we started from a uniform background plasma,
very soon (owing to the dense beam) strong density fluctuations appeared (see
Fig.~\ref{fig5}). During the evolution, their characteristic lengths became
longer (compare Figs.~\ref{fig5}a and \ref{fig5}b). In the early stages of
evolution, the associated electric field densities are not correlated with the
density fluctuations. On the other hand, in the later stages of evolution,
e.g., at $\omega_\mathrm{pe} t$ = 190 it can be seen that the electric fields
(Langmuir waves) start to be trapped in density depressions (see e.g. the
electric field densities at $z$, equal to $300\Delta$ and $430 \Delta$
respectively).

\begin{figure}[!t]
  \begin{center}
    \epsfig{file=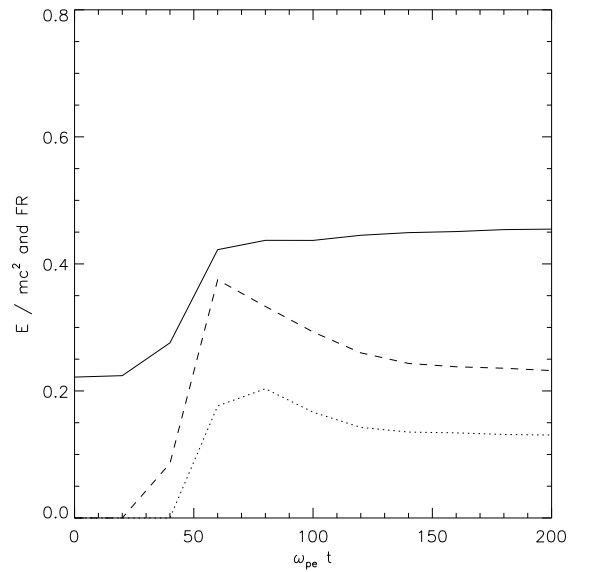, width=8cm}
        \end{center}
    \caption{Fractions of the beam energy in electrons
with energies greater than $E/mc^2$ = 0.25 (dashed line) and greater than
$E/mc^2$ = 0.3 (dotted line), for model E. The initial energy of the beam
electron is $E/mc^2 = 0.22$. For comparison, we plot the time evolution of the
maximum energy of the electron (solid line).}
  \label{fig6}
\end{figure}

To investigate the energetics of the accelerated electrons, we computed their
energy (expressed in fraction of the initial beam energy) above some selected
energy levels. The time evolution of these fractions for the energies higher
than $E/mc^2$ = 0.25 and $E/mc^2$ = 0.3, for model E are shown in
Fig.~\ref{fig6}. The initial energy of the beam electron was $E/mc^2$ =
0.22178. In this figure, for comparison the time evolution of the maximum
energy of the electron ($E/mc^2$) is plotted. The fractions firstly increase in
time, owing to the some time taken for acceleration of these electrons to the
selected energy levels. The fractions then slowly decrease to some saturation
level, in agreement with the saturation of the quasi-linear relaxation process.
The fractions generally decrease with the increase in the energy interval
between the beam energy and selected energy level. We also computed the
fractions of these accelerated electrons for the energy levels just above the
beam energy. For all computational models (A-H) at the time $\omega_\mathrm{pe}
t$ = 200, these fractions are summarized as percentages in the last column of
Table~1. These fractions increase as the magnetic field increase, e.g. from FR
= 10 $\%$ for $\omega_\mathrm{ce}/\omega_\mathrm{pe}$ = 0 (model A) to FR = 29
$\%$ for $\omega_\mathrm{ce}/\omega_\mathrm{pe}$ = 1 (model E). For even
greater values of $\omega_\mathrm{ce}/\omega_\mathrm{pe}$, the fractions start
to decrease (see FR = 27 $\%$ for $\omega_\mathrm{ce}/\omega_\mathrm{pe}$ = 1.3
in model F). Similar results were found for models G and H, where we considered
a beam density lower than in models A-F (see Table 1). As shown and analyzed in
the papers of \citet{2009NPGeo..16..525K} and \citet{2009A&A...506.1437K} for
the models with weak magnetic fields (models A-B), the Weibel instability is
significant and strongly influences the resulting electron distribution
function. On the other hand, the strong magnetic field (e.g. model E) reduces
the role of the Weibel instability (see also the discussion and
Fig.~\ref{fig8}).

As an illustration, we present in Fig.~\ref{fig4} the electron energy
distributions at $\omega_\mathrm{pe} t$ = 200 for all runs, i.e. for
$\omega_\mathrm{ce}/\omega_\mathrm{pe}$ = 0.0, 0.1, 0.5, 0.7, 1.0, and 1.3,
respectively (models A-F, Table 1). The number of electrons accelerated above
the initial electron-beam energy (expressed by the vertical dashed line)
corresponds to the fractions FR in Table 1.

\begin{figure}[!t]
  \begin{center}
    \epsfig{file=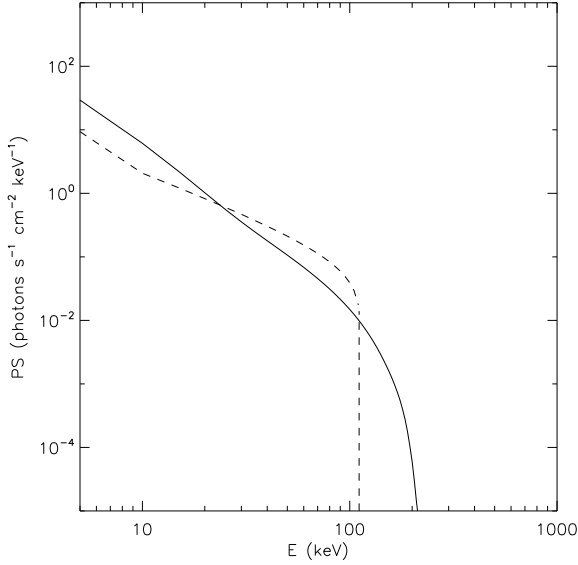, width=8cm}
        \end{center}
    \caption{The X-ray spectrum for model E at $\omega_\mathrm{pe}
t$ = 200 (solid line) and the initial state (dashed line).}
  \label{fig7}
\end{figure}

\begin{figure}[!t]
  \begin{center}
    \epsfig{file=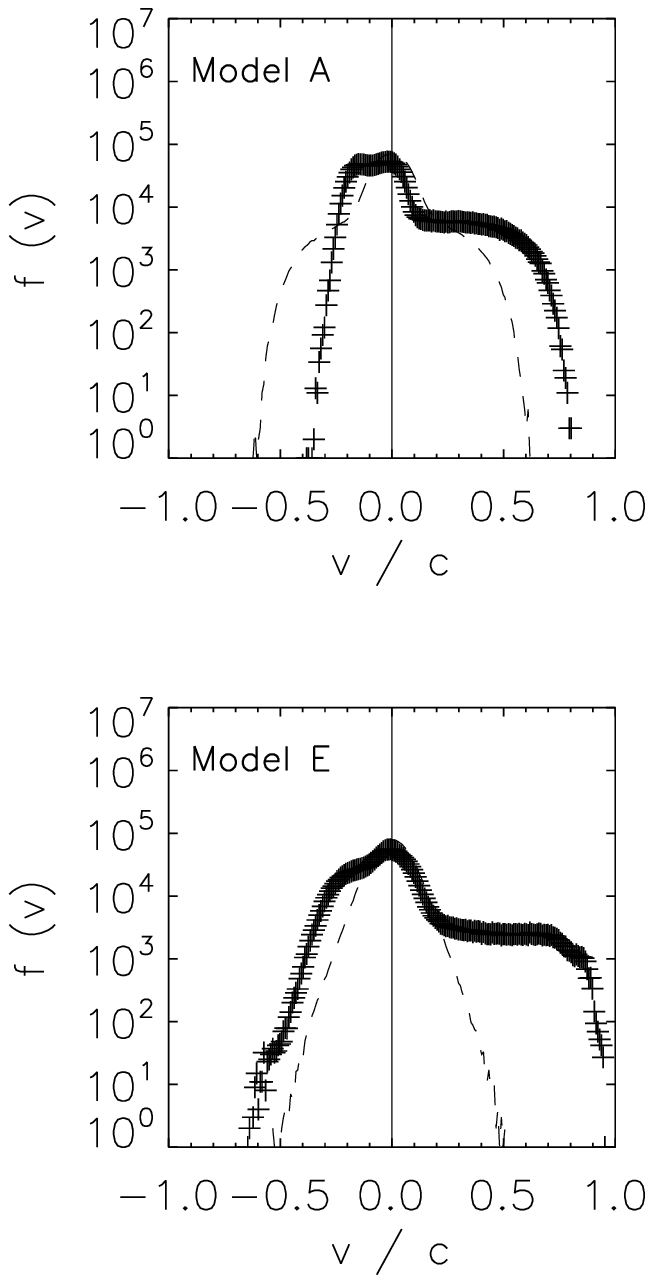, width=8cm}
        \end{center}
    \caption{Comparison of the electron distribution functions in model A (without
    the magnetic field) and model E ($\omega_\mathrm{ce}/\omega_\mathrm{pe}$ = 1.0).
     Crosses correspond to $f(v_{z})$ and dashed lines display overlapping $f(v_x)$ and
$f(v_y)$.}
  \label{fig8}
\end{figure}

\section{Quasi-linear relaxation -- analytical estimates}

While the analytical treatment of a 3D non-linear beam-plasma system is
impossible, the efficiency of acceleration can be estimated in the quasi-linear
limit. In this limit, the evolution of energetic electrons is described by two
coupled kinetic equations
\citep{1963JETP...16..682V,1967PlPh....9..719V,1995lnlp.book.....T}. For a
weakly magnetized plasma, when the field is strong enough to provide
one-dimensional electron dynamics, the electron beam evolution can be described
using standard quasi-linear theory
\begin{equation}
\frac{\partial f}{\partial t}= \frac{4\pi^2 e^2}{m^2}\frac{\partial}{\partial {v}}\left(
\frac{W_k}{{  v}}\frac{\partial f}{\partial { v}}\right) \;,
\label{eqk1}
\end{equation}
\begin{eqnarray}
\frac{\partial W_k}{\partial t}-\frac{\partial \omega_{pe}(x)}{\partial x}
\frac{\partial W_k}{\partial k} =\frac{\pi\omega_{pe}}{n_e} \frac{\omega_{pe}^2}{k^2}W_k\frac{\partial f}{\partial { v}} \;,
\label{eqk2}
\end{eqnarray}
where $f({\mr v},t)$ is the electron distribution function, $W_k$ is the
spectral energy density of the Langmuir waves, $\omega_{pe}$ is an electron
plasma density, and $e$ and $m$ are the electron charge and mass. The equations
describe the resonance ($\omega _{pe}=kv$) interaction of the electron beam
with the surrounding plasma via the generation of plasma waves and include
plasma inhomogeneity effects \citep[see,
e.g.][]{1967PlPh....9..719V,1969JETP...30..131R,2002PhRvE..65f6408K}.

When the plasma is uniform  $\frac{\partial \omega_{pe}(x)}{\partial x}=0$, the
stationary solution for the initially unstable beam distribution $ f({\mr v},
t=0)= g_0(v)$, $v<v_b$ of the coupled quasi-linear equations is well-known
\citep{1963JETP...16..682V}. The electron distribution function has the form of
a plateau
\begin{equation}
f({ v}, t\rightarrow \infty )= \frac{n _b}{v_b},  \;\;\;\;\; v<v_b,
\label{eq:f_t0}
\end{equation}
and the spectral energy density of plasma waves $W_k$ can be found from Equations (\ref{eqk1},\ref{eqk2})
using Equation  (\ref{eq:f_t0})
\begin{equation}
W(k=\frac{\omega _{pe}}{v}, t \rightarrow \infty)= \frac{m}{\omega _{pe}}
v^{3}\int_{0}^{v}\left(\frac{n_b}{v_b}-g_0(v)\right)\mbox{d}v, \; v<v_b,
\label{eq:W_t0}
\end{equation}
thus the spectral energy density becomes
\begin{equation}
W(k=\frac{\omega _{pe}}{v}, t \rightarrow \infty)= \frac{m n_b}{\omega _{pe}v_b} v^{4}, \;\; v<v_b
\label{eq:W_t0_1}
\end{equation}
for $g_0(v)=n_b\delta (v-v_b)$ and takes the form
\begin{equation}
W(k=\frac{\omega _{pe}}{v}, t \rightarrow \infty)= \frac{m n_b}{\omega _{pe}v_b} v^{3}\left(1-\frac{v}{v_b}\right), \;\; v<v_b
\label{eq:W_t0_2}
\end{equation}
for $g_0(v)=2n_bv/v_b^2$, $v<v_b$, \citep[see][for
details]{2001A&A...375..629K}. The total energy density of plasma waves from
Equation (\ref{eq:W_t0_1}) is $U_{L}=2/3 \times mn_bv_b^2/2$ and $U_{L}=1/3
\times mn_bv_b^2/2$ for the second case given by Equation (\ref{eq:W_t0_2}).

However, when the plasma has a positive density gradient $\frac{\partial
\omega_{pe}(x)}{\partial x}>0$, which corresponds to electrons propagating into
the region of higher density (as in the standard flare scenario), the Langmuir
waves slowly evolve towards larger phase velocities or smaller wavenumbers $k$.
As Langmuir wave packets propagate within the plasma, the total energy of the
wave packet $\omega (k,x)$ must be constant \citep{1967PlPh....9..719V}, which
requires that the wavenumber change be negative for a positive plasma-density
gradient, such that
\begin{equation}
{\Delta k}\simeq -\frac{\partial \omega _{pe}(x)}{\partial x}\Delta t.
\label{eq:delta_k}
\end{equation}
Therefore, the motion of Langmuir wave over the time range $\Delta t$ results
in a decrease in the wavenumber from $k$ to $k-\frac{\partial \omega
_{pe}(x)}{\partial x}\Delta t$. The waves shifted to higher $v$ (smaller $k$)
can be effectively re-absorbed by the beam, which leads to an acceleration of
the electrons and the formation of an extended plateau above $v>v_b$. Using the
conservation of energy, and assuming that all Langmuir waves are re-absorbed by
the beam owing to the plasma inhomogeneity,  $W(k=\omega/{v}, t \rightarrow
\infty)= 0$, one finds that for the energy
\begin{equation}
\int_0^{\infty} f({v}, t\rightarrow \infty )v^2 dv= \int_0^{\infty} f({v}, t=0)v^2 dv
\label{eq:energy}
\end{equation}
the electron distribution has electrons with velocity $v>v_{b}$
\begin{equation}
f({v}, t\rightarrow \infty )= \frac{n_b}{v_{max}},  \;\;\;\;\; v<v_{max},
\label{eq:f}
\end{equation}
where the new maximum velocity is from Equation (\ref{eq:energy})
\begin{equation}
v_{max} ^2= \frac{3}{n_b} \int_{0}^{\infty} g_0(v)v^2dv.
\label{eq:v_max}
\end{equation}
Numerical solutions of equations  (\ref{eqk1}-\ref{eqk2}) \citep[see][for
details]{2001A&A...375..629K} show that the value of the density gradient
mostly affects the rate of the extended plateau formation and that the final
state of the initially unstable distribution $g_0(v)=2n_bv/v_b^2$, $v<v_b$,
gives $v_{max}=\sqrt{3/2} v_b$.  In the case of $g_0(v)=n_b\delta (v-v_b)$, one
finds that $v_{max}=\sqrt{3} v_b$.

In the case of non-linear interactions and various density fluctuations, the
energy exchange between the electrons and plasma waves becomes more
complicated. \citet{2012A&A...539A..43K} used numerical simulations to estimate
the role of these effects. Here, assuming that small-scale density fluctuations
in the plasma generated during the beam-plasma instability are random with zero
mean, the evolution of Langmuir waves in plasma can be approximated as a
symmetric diffusion in $k$-space \citep[see Eq 14 in][]{2012A&A...539A..43K}.
This leads to equal numbers of plasma waves spreading towards smaller and
larger the phase velocities. The former waves will be Landau-absorbed by the
thermal plasma, while the latter will be absorbed by the beam resulting in
electron acceleration in the tail of distribution. Therefore, for the Langmuir
wave spectrum flattened in $k$-space, (see Figure \ref{fig2}), these simplistic
arguments suggest that only half of the Langmuir wave energy given by Equations
(\ref{eq:W_t0_1}, \ref{eq:W_t0_2}) will be reabsorbed back. The maximum
velocity then becomes $v_{max}=\sqrt{2} v_b$ and the energy of the electrons
with $v>v_b$ becomes
\begin{eqnarray}
{U({v}> {v}_b, t\rightarrow \infty )}= \frac{m}{2}\int _{v_b}^{v_{max}}\frac{n_b}{v_{max}}v^2dv \cr
 \simeq 0.43 \frac{m n_b v_b^2}{2},
\label{eq:E_max1}
\end{eqnarray}
for $g_0(v)=n_b\delta (v-v_b)$. We note that these estimates are rather close
to the numbers inferred from the 3D PIC  simulations presented in Table 1. It
is worth noting that the collisional relaxation of the electron power-law
spectrum, which is initially stable was considered by
\citet{2012A&A...539A..43K}, while in this paper we treat a `classical' case of
beam-plasma instability.

\section{Discussion and conclusions}

We have performed a number of 3D PIC simulations of the beam-plasma instability
with monoenergetic beams and have shown that during relaxation a population of
electrons with velocities exceeding those of the injected electrons appears.
The energy of these electrons is around $10-30\%$ of the initial beam energy.

Using PIC simulations, it is difficult to predict the long-term time evolution
of these processes in solar flares. However, as shown in Fig.~\ref{fig7} this
effect is indicated by the high-energy limit of the X-ray spectrum, namely that
the accelerated high-energy electrons shift the X-ray spectrum to higher
energies. This result together with the radio diagnostics can be used for an
estimation of these acceleration processes. For example, if we take the
dm-spikes as a radio signature of the acceleration process in solar flares
\citep{1991A&A...251..285G}, then the advanced theory of these bursts can be
used to estimate the electron distribution function at the acceleration site.
Comparing this function with that determined from the hard X-ray spectrum at
the flare footpoints, the acceleration efficiency can then be estimated.

We have found that the increasing magnetic field strength leads to a larger
fraction of accelerated electrons. Therefore, we decided to compare the
electron distribution functions of the cases with and without a magnetic field
(in models A and E, Fig.~\ref{fig8}). As presented and analyzed in
\citet{2009NPGeo..16..525K} and \citet{2009A&A...506.1437K}, the main
difference in both cases is caused by the Weibel instability. In model E, the
Weibel instability is reduced, while in model A (without the magnetic field)
the Weibel instability transfers the beam energy to a heating of mainly
perpendicular components of the background plasma. Thus, in the case without
the magnetic field, not only the bump-on-tail instability but also the Weibel
instability operates and less energy (than in model E) is transferred to the
Langmuir waves, leading to a weaker acceleration.

We considered two values of the ratio of the beam to background plasma
densities $n_\mathrm{b}/n_\mathrm{e}$ = 1/8 and 1/40, which imply that the
return-current electron speeds are $v_\mathrm{d}$ = 0.083 c and $v_\mathrm{d}$
= 0.016 c (where c is the speed of light), respectively. This corresponds to
two regimes of the return-current electron speed either greater or lower than
the thermal plasma velocity, which is in our model $v_{T\mathrm{e}}$ = 0.06
$c$. In both cases, the percentage of the energy in accelerated electrons is
similar. The case where $n_\mathrm{b}/n_\mathrm{e}$ = 1/8, i.e. the case with
the drift speed greater than the initial thermal velocity, should be unstable
for the Buneman instability. However, the time of evolution considered in the
present study is shorter than the time required to develop such an instability.
For much longer term evolutions of similar systems in this regime computed with
a 1-D Vlasov code, the formation of weak double layers was found and proposed
as an explanation of broken power-law X-ray spectra in solar flares by
\citet{2008A&A...478..889L} (see also Karlick\'y 2012).

We note that the size of our numerical box
(45$\Delta$$\times$45$\Delta$$\times$600$\Delta$) is limited by the memory and
speed limits of our computer. The limited number of spatial grids limits a
number of grids in the $k$-vector space. This limitation can influence the
wave-wave and wave-particle interactions in this acceleration process. This is
especially important in the linear regime of these interactions, which would
correspond here to the cases with low density beams. However, in the present
study we considered very dense beams ($n_b/n_e$ = 1/8 and 1/40), which generate
strong density fluctuations ensuring that these interactions are far from
linear regimes (We note that in the interpretation of the hard X-ray emission
the dense electron beams are often required).  We made additional tests for
model E. We used the same parameters as in model E, but we varied the size of
the numerical system. We obtained similar results. For the system size
(20$\Delta$$\times$20$\Delta$$\times$1500$\Delta$) at the time
$\omega_\mathrm{pe} t$ = 200, the fraction FR is 26 $\%$ and for the size
(10$\Delta$$\times$10$\Delta$$\times$3000$\Delta$) at the same time the
fraction FR = 25 $\%$; compare with the fraction FR = 29 $\%$ for model E in
Table 1. Nevertheless, to make these results more precise we plan to repeat
these computations in larger numerical boxes on a more powerful computer.

Comparing the numerical and analytical treatment, we propose that for this type
of acceleration the density fluctuations and non-linear wave-wave interactions
are essential. While in a strictly uniform plasma this acceleration is
impossible (see the analytical estimations), in real conditions with
sufficiently dense electron beams some beam electrons are accelerated to
energies greater than the initial ones via non-linear wave interactions and
density fluctuations.

In solar flares, high non-thermal electron fluxes are often required to explain
the observed X-ray emission, which provide the suitable conditions for fast
beam-plasma instability. As we have shown, the beam-plasma instability
generates the Langmuir wave turbulence. The acceleration of electrons occurs
owing to the $k$-space evolution of Langmuir waves, in which not only the
Langmuir waves with phase velocities smaller than the initial beam velocity are
generated (as in the case of beam relaxation in uniform plasma) but also those
with higher phase velocities. The process of electron acceleration is fast, and
occurs on a timescale much shorter than the electron transport time from an
acceleration site to the dense chromospheric region. We therefore expect that
during these processes the tail of the distribution extends towards higher
energies. The number of high-energy electrons above some energy level increases
and thus influences the hard X-ray spectrum. However, the total number of beam
electrons is conserved and the process only redistributes the beam energy.
These results emphasize that the transport of electrons should not be treated
using a single particle description, and that the collective effects produce
not only energy loss, but an effective acceleration of electrons. The analysis
of a hard X-ray spectrum assuming only collisional losses could therefore infer
that a larger number of electrons than actually is initially accelerated owing
to the additional in-flight acceleration.

\begin{acknowledgements}
The authors thank the anonymous referee for comments that improved the paper.
All computations were performed on the parallel computer OCAS (Ond\v{r}ejov
Cluster for Astrophysical Simulations, see http://wave.asu.cas.cz/ocas). This
research was supported by the grant P209/12/0103 (GA CR). Financial support by
the STFC rolling grant (EPK), the Leverhulme Trust (EPK), and the European
Commission through the HESPE (FP7-SPACE-2010-263086) (EPK) and "Radiosun"
(PEOPLE-2011-IRSES-295272) Networks (MK, EPK) is gratefully acknowledged.
\end{acknowledgements}

\bibliographystyle{aa}
\bibliography{19400refs}

\end{document}